\documentclass[
amsmath, %
floatfix, %
twocolumn, %
reprint, %
superscriptaddress, %
apl, %
aip, %
citeautoscript, %
]{revtex4-1}
\usepackage[T1]{fontenc}
\usepackage[utf8]{inputenc}
\usepackage{graphicx}
\usepackage{latexsym}
\usepackage{amsthm}

\usepackage{color}
\usepackage{mathtools}
\usepackage{mathdots}
\usepackage{xcolor}
\usepackage[low-sup]{subdepth}

\usepackage[]{newtxtext}
\usepackage[subscriptcorrection,nosymbolsc,smallerops,bigdelims]{newtxmath}
\DeclareMathAlphabet{\mc}{OMS}{cmsy}{m}{n}

\DeclareMathAlphabet{\mcb}{OMS}{cmsy}{b}{n}
\usepackage{bm}

\newcommand{\widebar}[1]{#1^{\bm{-}}}

\usepackage{floatrow}
\mathchardef\mhyp="2D
\usepackage{siunitx}
\usepackage{hyperref}
\hypersetup{
	colorlinks,
	linkcolor={blue!80!black},
	citecolor={blue!80!black},
	urlcolor ={blue!80!black}
}
\makeatletter
\renewcommand*{\eqref}[1]{%
	\hyperref[{#1}]{\textup{\tagform@{\ref*{#1}}}}%
}
\makeatother

\DeclarePairedDelimiter\lr{\lparen}{\rparen}
\DeclarePairedDelimiter\Lr{\lbrack}{\rbrack}

\DeclarePairedDelimiterX{\comm}[2]{\lbrack}{\rbrack}{#1, #2}

\DeclarePairedDelimiterX{\braket}[2]{\langle}{\rangle}{#1\delimsize\vert #2}
\DeclarePairedDelimiterX{\ketbra}[2]{\rvert}{\lvert}{#1 \delimsize\rangle\!\delimsize\langle #2}
\DeclarePairedDelimiterX{\matrixel}[3]{\langle}{\rangle}{#1 \delimsize\vert\,\mathopen{}#2\,\delimsize\vert\,\mathopen{} #3}

\newcommand{\upcapt}{\vspace{-0.2em}}
\newcommand{\upcaptb}{\vspace{-1em}}

\newcommand{\mr}[1]{\mathrm{#1}}

\newcommand{\eqnref}[1]{Eq.~\kern-0.5pt\eqref{#1}}
\newcommand{\refref}[1]{Ref.~\kern-0.5pt[\onlinecite{#1}]}
\newcommand{\figref}[1]{Fig.~\kern-0.5pt\ref{#1}}

\newcommand{\secref}[1]{Sec.~\kern-0.5pt\ref{#1}}
\newcommand{\subeqnref}[2]{Eq.~\kern-0.5pt\hyperref[#1]{(\ref*{#1}#2)}}
\newcommand{\subfigref}[2]{Fig.~\kern-0.5pt\hyperref[#1]{\ref*{#1}(#2)}}

\newcommand{\ex}{{{}^{{*}}}}

\newcommand{\Xm}{\widebar{\mathrm{X}}}
\newcommand{\XmE}{\widebar{\mathrm{X}}\ex}
\newcommand{\XXm}{\widebar{\mathrm{XX}}}

\begin{document}
	\title{Strongly temperature-dependent recombination kinetics of a negatively charged exciton in asymmetric quantum dots
		at 1.55 \si{\micro}m}
	
	\author{{\L}. Dusanowski}
		\affiliation{Laboratory for Optical Spectroscopy of Nanostructures, Department of Experimental Physics, Faculty of Fundamental Problems of Technology, Wroc\l{}aw University of Science and Technology, 50-370 Wroc\l{}aw, Poland}
		\affiliation{Technische Physik, Universit\"at W\"urzburg and Wilhelm Conrad R\"ontgen Research Center for Complex Material Systems, Am Hubland, D-97074 W\"urzburg, Germany}
			
	\author{M. Gawe{\l}czyk}
		\email{michal.gawelczyk@pwr.edu.pl}
		\affiliation{Laboratory for Optical Spectroscopy of Nanostructures, Department of Experimental Physics, Faculty of Fundamental Problems of Technology, Wroc\l{}aw University of Science and Technology, 50-370 Wroc\l{}aw, Poland}
		\affiliation{Department of Theoretical Physics, Faculty of Fundamental Problems of Technology, Wroc\l{}aw University of Science and Technology, 50-370 Wroc\l{}aw, Poland}
		
	\author{J. Misiewicz}
		\affiliation{Laboratory for Optical Spectroscopy of Nanostructures, Department of Experimental Physics, Faculty of Fundamental Problems of Technology, Wroc\l{}aw University of Science and Technology, 50-370 Wroc\l{}aw, Poland}

	\author{S. H\"{o}fling}
		\affiliation{Technische Physik, Universit\"at W\"urzburg and Wilhelm Conrad R\"ontgen Research Center for Complex Material Systems, Am Hubland, D-97074 W\"urzburg, Germany}
		\affiliation{SUPA, School of Physics and Astronomy, University of St.\ Andrews, North Haugh, KY16 9SS St.\ Andrews, United Kingdom}

	\author{J. P. Reithmaier}
		\affiliation{Institute of Nanostructure Technologies and Analytics (INA), CINSaT, University of Kassel, Heinrich-Plett-Str. 40, 34132 Kassel, Germany}
		
	\author{G. S\k{e}k}
		\affiliation{Laboratory for Optical Spectroscopy of Nanostructures, Department of Experimental Physics, Faculty of Fundamental Problems of Technology, Wroc\l{}aw University of Science and Technology, 50-370 Wroc\l{}aw, Poland}

	\begin{abstract}
		 We report on strongly temperature-dependent kinetics of negatively charged carrier complexes in asymmetric InAs/AlGaInAs/InP quantum dots (dashes) emitting at telecom wavelengths. The structures are highly elongated and of large volume, which results in atypical carrier confinement characteristics with $s$-$p$ shell energy splittings far below the optical phonon energy, which strongly affects the phonon-assisted relaxation. Probing the emission kinetics with time-resolved microphotoluminescence from a single dot, we observe a strongly non-monotonic temperature dependence of the charged exciton lifetime. Using a kinetic rate-equation model, we find that a relaxation side-path through the excited charged exciton triplet states may lead to such behavior. This, however, involves efficient singlet-triplet relaxation via the electron spin-flip. Thus, we interpret the results as an indirect observation of strongly enhanced electron spin relaxation without magnetic field, possibly resulting from atypical confinement characteristics.
	\end{abstract}
	
	\maketitle

	Due to the three-dimensional quantum confinement, various carrier complexes can be created within semiconductor nanostructures, including charged excitons (trions) that are composed of the electron-hole pair and an additional electron or hole. Trions have been extensively studied in quantum dots (QDs) from the point of view of resident spin initialization\cite{KuglerPRB2011,GawelczykPRB2013}, single photon generation\cite{StraufNaturePhot2007} (no dark states limiting emission rates, contrary to neutral excitons) as well as in implementations of light-matter interfaces for spin-photon\cite{GaoNature2012,DeGreveNature2012,DeGreveNatureComms2013} and spin-spin entanglement generation.\cite{DelteilNaturePhys2015,DelteilPRL2017,StockillPRL2017,KochPSSc2006,KazimierczukPRB2010} For these applications, understanding the occupation kinetics of carrier states and the resulting light emission is of particular importance. While it has been studied at low temperatures,\cite{PattonPRB2003,PoemPRB20010,Munoz-MatutanoPhysicaE2008,IgarashiPRB2010} systematic studies concerning the temperature dependence of trion emission dynamics in epitaxial nanostructures are needed. 

	Quantum dashes (QDashes) are nanostructures that are strongly elongated in one of the in-plane directions, with discrete carrier energy levels inherited from similar but much more symmetric QDs\cite{UtzmeierAPL1996,GuoAPL1997}. Typically, their width is in the range of $\SIrange{10}{30}{\nano\metre}$, whereas the length may vary up to above \SI{100}{\nano\metre}. In InAs on InP structures, this shape asymmetry may emerge spontaneously during the epitaxial growth in a self-assembled manner.\cite{SauerwaldAPL2005,BraultAPL1998} The optical properties of such QDashes have been the subject of many studies focused on both ensembles\cite{DeryJAP2004,ReithmaierJPhysD2008,CapuaOptExpress2012,SyperekAPL2013} and single objects.\cite{SekJAP2009,DusanowskiAPL2013,DusanowskiAPL2014,DusanowskiPRB2014,MrowinskiAPL2015,MrowinskiPRB2016,DusanowskiAPL2016} This has been motivated by their compatibility with modern semiconductor-based optoelectronics and nanophotonics, as well as the emission at telecom wavelengths. This makes such structures promising for future technologies involving fiber-based quantum communication and information processing. Recently, single-photon emission from individual QDashes at telecom wavelengths up to $T\mathbin{=}\SI{80}{\kelvin}$ has been demonstrated,\cite{DusanowskiAPL2016} showing their suitability for implementation in quantum-information technologies.
	
	Here, we investigate the dynamics of emission from charged excitons, focusing on effects related to the kinetics of transitions between ground and excited states of a negatively charged trion. In QDashes, the single-particle level spacings for both electrons and holes are significantly below the optical-phonon energy\cite{SyperekAPL2013,GawelczykPRB2017}, so relaxation processes mediated by acoustic phonons, leading to temperature-dependent occupation kinetics, are expected. We show that the dynamics of the trion recombination cascade in QDashes exhibits an uncommon, non-monotonic temperature dependence, which we probe by measuring the trion emission lifetime as a function of temperature. Using a rate-equation model including the resident electron, ground and excited trion as well as charged biexciton states, we are able to qualitatively reproduce the observed dependence. However, this is only possible if efficient electron spin relaxation leading to transitions among excited trion states is present. The observed nonmonotonicity is therefore understood by us as originating from occupation transfer along a side-path trough excited-trion triplet states.
	
	The investigated sample was grown in a gas-source molecular beam epitaxy system on an S-doped InP(001) substrate. The layer sequence begins with a 200~nm thick Al$_{0.24}$Ga$_{0.23}$In$_{0.53}$As barrier layer lattice matched to InP, which was grown on the substrate at \SI{500}{\celsius}. To form QDs in a self-assembled way, an InAs layer with the nominal thickness of \SI{1.3}{\nano\meter} was deposited at \SI{470}{\celsius}, from which nanostructures on a wetting layer were formed. Next, QDs were covered with a \SI{100}{\nano\meter} thick barrier layer, and subsequently capped with \SI{10}{\nano\meter} of InP. Due to the anisotropy of the surface diffusion coefficient, the dots are strongly elongated in shape, preferentially in the $[1\bar{1}0]$ direction, with small random deviations\cite{BraultAPL1998}. The typical dimensions of QDs under study are about \SI{20}{\nano\meter} in width and \SI{\sim 3.5}{\nano\meter} in height, while their length possibly varies from tens to hundreds of nanometers.\cite{SauerwaldAPL2005, GawelczykPRB2017} Since the planar density of structures is rather high, above \SI{e10}{\centi\meter^{-2}}, a combination of electron beam lithography and etching techniques was used to produce mesas of various sizes down to \SI{0.125}{\micro\meter^2} in order to resolve the emission from individual objects of the inhomogeneous ensemble. A weak residual n-type doping is possible in these structures.
			
	For all the experiments, the sample was kept in a liquid-helium continuous-flow cryostat equipped with a heating wire attached to a PID temperature regulator loop. The microphotolumienscence (\si{\micro}PL) and time-resolved \si{\micro}PL (\si{\micro}TRPL) measurements were performed using a setup that provides a spatial resolution of the order of single \si{\micro}m, and a spectral resolution of \SI{\sim 100}{\micro\electronvolt}. For the \si{\micro}PL measurements the sample was excited with the \SI{787}{\nano\meter} line of a continuous-wave (CW) laser. The intensity of emission into isolated spectral lines was detected by a liquid-nitrogen-cooled GaInAs-based linear detector combined with a \SI{0.3}{\meter} focal length single grating monochromator. For \si{\micro}TRPL experiments, the structure was excited non-resonantly with a train of \SI{160}{\femto\second} long pulses generated by a mode-locked Ti:Sapphire laser at a repetition frequency of \SI{76}{\mega\hertz} and the photon energy of \SI{1.49}{\electronvolt} (\SI{830}{\nano\metre}). The \si{\micro}TRPL signal was filtered by a monochromator and photons were collected by an NbN superconducting nanowire detector with a temporal response of \SI{50}{\pico\second} (the overall setup resolution is  \SI{\sim80}{\pico\second}). The \si{\micro}TRPL was measured using the time-correlated single-photon-counting method. A multichannel event timer was synchronized with the pulse laser to obtain photon-event statistics.

	\begin{figure}[tb]
		\includegraphics[width=\columnwidth]{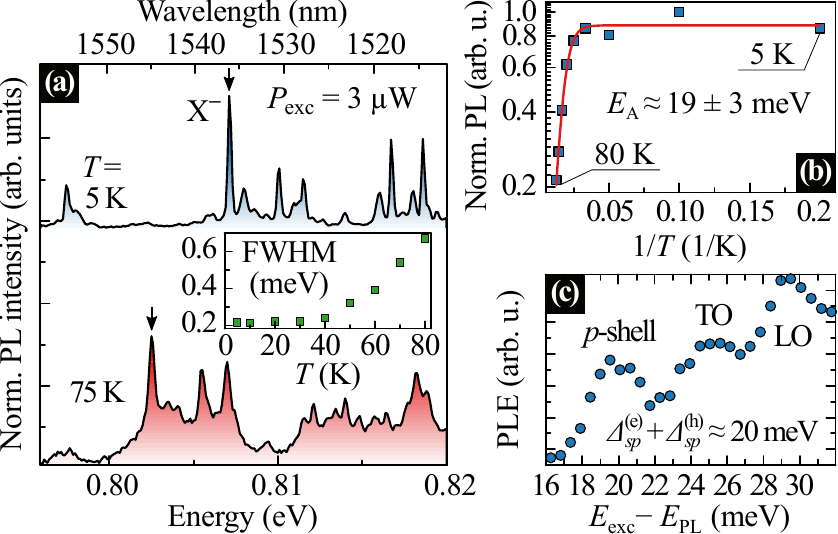}\upcapt
		\caption{\label{fig:micropl}(a)~Normalized \si{\micro}PL spectra from a mesa with QDashes recorded at $T\mathbin{=}\SI{5}{\kelvin}$ and $\SI{75}{\kelvin}$, at an average excitation power of \SI{3}{\micro\watt}; inset: temperature dependence of the $\Xm$ line full width at half-maximum. (b)~Arrhenius plot of the $\Xm$ line PL intensity with a fit (solid line). (c)~PLE spectrum from a bigger ensemble ($2\mathbin{\times}2$\,\si{\micro\meter^2} mesa).\upcaptb}
	\end{figure}
	In \subfigref{fig:micropl}{a}, we present normalized (to maximum intensities) \si{\micro}PL spectra from a single QDash within the $500 \mathbin{\times} 250$\,\si{\nano\meter^2} mesa recorded at \SI{5}{\kelvin} and \SI{75}{\kelvin}. The spectral line marked with an arrow is the one identified as originating from the negatively charged exciton ($\Xm$), based on the characteristic linear excitation-power dependence of intensity and no fine-structure splitting as shown in previous studies\cite{DusanowskiAPL2014,DusanowskiAPL2016,MrowinskiPRB2016}. A common redshift of the luminescence peaks with temperature is present due to the thermal shrinkage of the material band-gap. At $T\mathbin{=}\SI{5}{\kelvin}$ the considered peak has a Gaussian profile with the linewidth of \SI{\sim 0.2}{\milli\electronvolt}, which indicates an inhomogeneous impact of the environment.\cite{SeufertAPL2000} At $T\mathbin{=}\SI{75}{\kelvin}$, luminescence side-bands appear likely due to the coupling of carriers to acoustic phonons.\cite{DusanowskiPRB2014} The inset presents the temperature dependence of the $\Xm$ emission line full width at half-maximum (FWHM), exhibiting an increase up to \SI{0.67}{\milli\electronvolt} at \SI{80}{\kelvin}. This resembles the behavior predicted for excitons in similar structures as resulting from the carrier-phonon coupling.\cite{DusanowskiPRB2014} In \subfigref{fig:micropl}{b}, we present an Arrhenius plot of the \si{\micro}PL $\Xm$ line intensity, where a five-fold drop is observed for $T\mathbin{=}\SIrange{5}{80}{\kelvin}$. Fitting of a $I\mathbin{=}I_0/\Lr{1+a\exp\lr{-E_\mr{A}/kT}}$ curve\cite{LerouxJAP1999} (solid line) to the data brought activation energy $E_\mr{A}\mathbin{\approx}\SI{19}{\milli\electronvolt}$. However, it should be noted that for multiple processes with various energies, such fitting yields an overestimated value for the lowest-energy process. Additionally, in \subfigref{fig:micropl}{c} we plot a photoluminescence excitation (PLE) spectrum collected from a bigger mesa ($2\mathbin{\times}2$\,\si{\micro\meter^2}), which allowed us to identify the $p$-shell bright state at $\SI{\sim20}{\milli\electronvolt}$ above the ground state. Based on the confinement characteristics,\cite{GawelczykPRB2017} this may be split into $\SI{\sim15}{\milli\electronvolt}$ and \SI{\sim5}{\milli\electronvolt} for electrons and holes, respectively.

	\begin{figure}[tb]
		\includegraphics[width=\columnwidth]{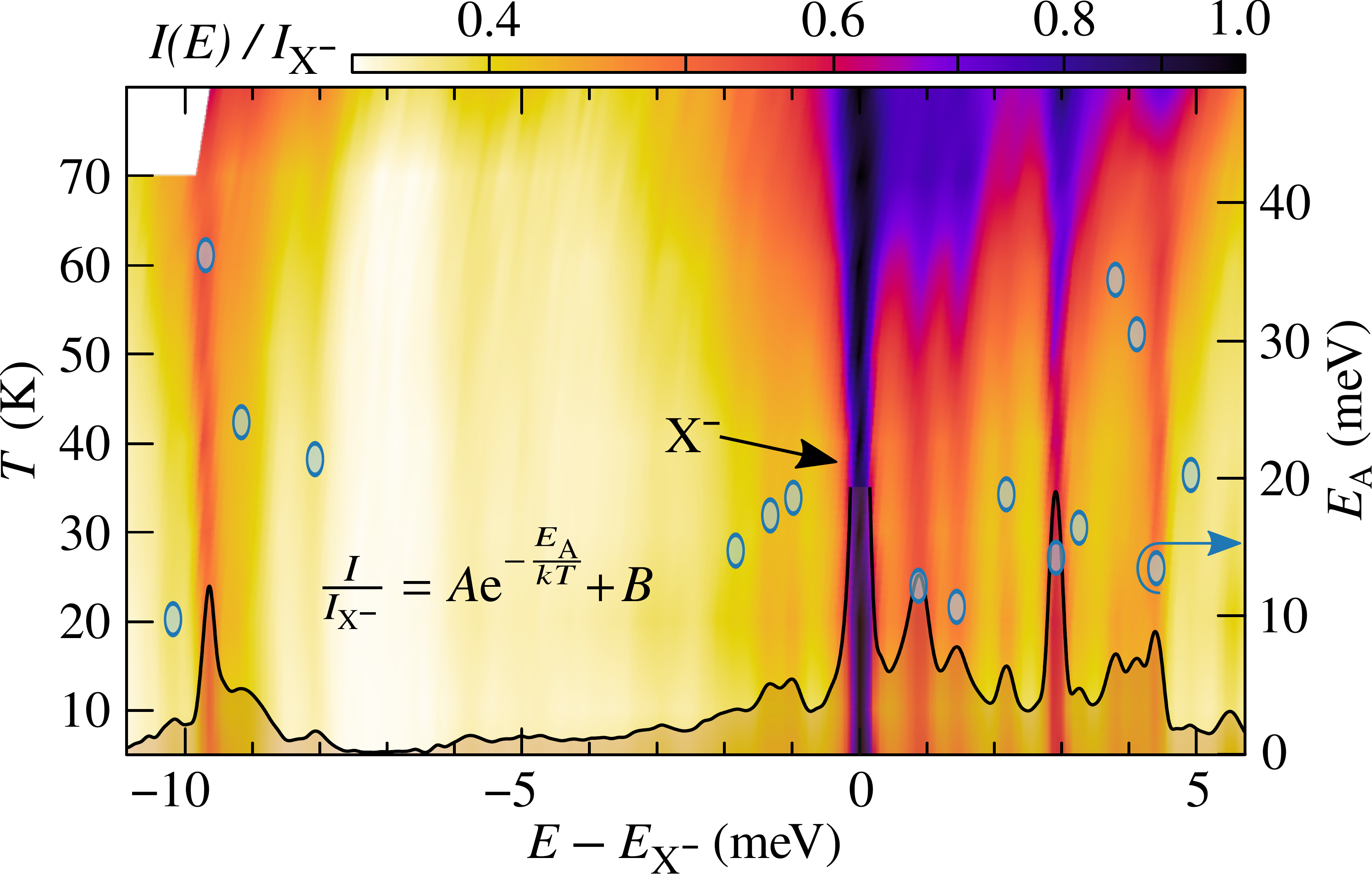}\upcapt
		\caption{\label{fig:micropl_thermal}Interpolated dependence of the \si{\micro}PL intensity $I$ relative to the $\Xm$ line intensity $I_{\Xm}$ (color map) on energy detuning from $\Xm$, $E-E_{\Xm}$, and temperature. Blue symbols (right axis) show activation energies obtained at peaks' spectral positions by fitting the given formula.\upcaptb}
	\end{figure}
	For a thermal escape of carriers\cite{PodemskiAPL2006} from a large QDash one may expect energies of $\SI{\sim300}{\milli\electronvolt}$ and $\SI{\sim50}{\milli\electronvolt}$ for ground-state electrons and holes, respectively.\cite{GawelczykPRB2017} The observed lower activation energy is possibly connected with the complex kinetics within the recombination cascade, since the value corresponds well to the estimated electron $s$-$p$ splitting.\cite{GawelczykPRB2017} In the presence of such thermally activated transitions, the emission may take place partly from states involving one excited electron ($\XmE$; at different wavelength), which reduces the intensity of the $\Xm$ line. 
	All absolute intensities exhibit thermal-quenching, most possibly due to the thermal escape of holes. To focus on lower-energy transitions among quantized levels, we present in Fig.~\ref{fig:micropl_thermal} a color map of intensities relative to the $\Xm$ line intensity as a function of detuning from $\Xm$ (to get rid of the redshift) and temperature. Assuming an approximately common nature of the carrier-escape-related quenching, these relative values should be free of it and give an insight into the occupation kinetics of the trion levels. Indeed, lines in the vicinity of $\Xm$ exhibit relative exponential thermal growth, corresponding activation energies $E_\mr{A}$ are plotted with the blue ovals. There is a number of lines with $E_\mr{A}\mathbin{=}\SIrange{10}{20}{\milli\electronvolt}$, which we consider to be a fingerprint of emission from singlet and triplet states of $\XmE$ ($E_\mr{S{/}T}\mathbin{\simeq}\varDelta_{sp}\pm\varDelta_\mr{ee}/2$). However, the intensities of these lines are too low to allow for a reliable time-resolved study and accurate identification.

	\begin{figure}[tb]
		\includegraphics[width=\columnwidth]{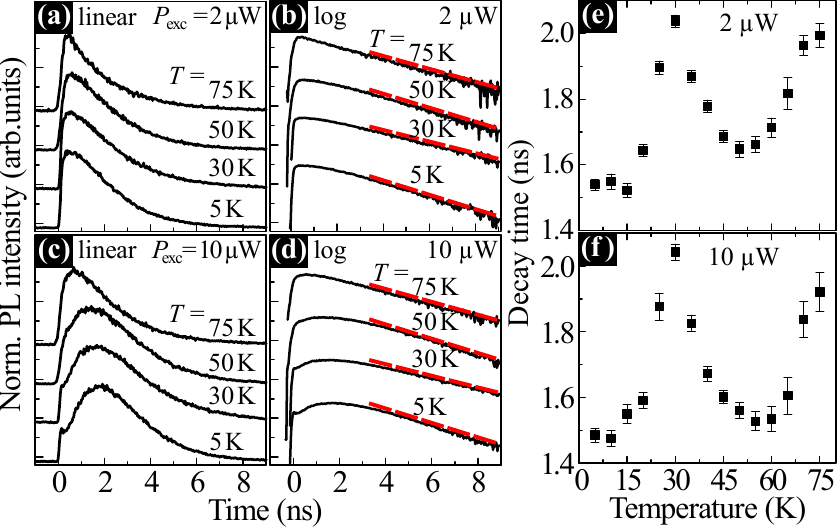}\upcapt
		\caption{\label{fig:trpl}(a), (b) PL time traces after a pulsed non-resonant excitation with the pulse average power $P\mathbin{=}\SI{2}{\micro\watt}$ and \SI{10}{\micro\watt}, respectively, at various temperatures. (c), (d) The same in the log scale with exponential fits (dashed red lines; vertically shifted for visibility). (e), (f) Temperature dependence of the PL decay time for $P\mathbin{=}\SI{2}{\micro\watt}$ and \SI{10}{\micro\watt}, respectively.\upcaptb}
	\end{figure}
	\begin{figure}[tb]
		\begin{center}
			\includegraphics[width=\columnwidth]{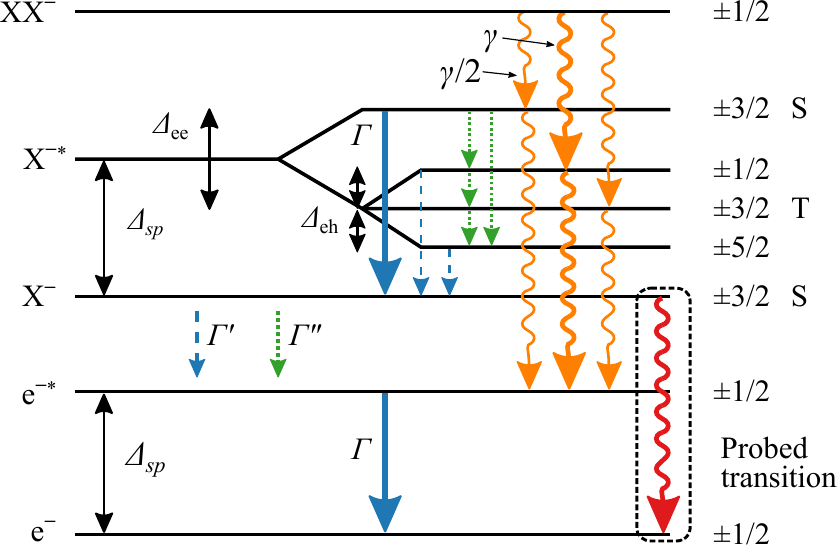}\vspace{-0.5em}
		\end{center}
		\caption{\label{fig:ladder}Schematic energy ladder of states included in the rate-equation model: the ground (excited) electron state $\mr{e}^{\bm{-}(*)}$, ground (excited) trion $\mr{X}^{\bm{-}(*)}$, and negatively charged biexciton $\mr{XX}^{\bm{-}}$. Energy splittings and transitions are marked: radiative transitions (wavy arrows, thickness corresponds to the oscillator strength\cite{AkimovPRB2005}), spin-preserving relaxation (solid arrows), orbital relaxation with a spin flip (dashed arrows), spin-flips (dotted arrows).\upcaptb}
	\end{figure}
	In \subfigref{fig:trpl}{a--d}, we present the results of TRPL experiments on the $\Xm$ line recorded at $T\mathbin{=}\SIrange{5}{80}{\kelvin}$ for two average excitation powers of \SI{2}{\micro\watt} (top panels) and \SI{10}{\micro\watt} (bottom panels).
	Each of the low-temperature \si{\micro}TRPL traces shows the temporal-resolution-limited rise followed by a decay, time constant of which initially increases from \SI{~1.55}{\nano\second} up to \SI{~2}{\nano\second} for $\SIrange{5}{30}{\kelvin}$, then decreases to \SI{~1.7}{\nano\second} at \SI{55}{\kelvin}, to finally increase again up to \SI{~2}{\nano\second} at \SI{80}{\kelvin}. These values are obtained from fitting of an exponential function (dashed red lines) to the post-rise tails of the \si{\micro}PL decay traces to catch the long-time behavior, as a complex character of decays is expected and an experimentally-feasible measure is needed to compare with results from the rate-equation model. The resulting values are plotted in \subfigref{fig:trpl}{e}. The character of \si{\micro}PL decays obtained at the higher average excitation power is more complex. A double build-up of intensity may be found in the low-temperature traces in this case. The first one is trivial, due to the temporal resolution of the system, while the second, with the time constant of \SI{\sim 0.8}{\nano\second} may be associated with some kind of $\Xm$ refilling process. The latter may result either from the slow relaxation from the higher-energy states and/or from the charged biexciton ($\XXm$) radiative transition to the excited trion ($\XmE$) states (top three wavy orange arrows in Fig.~\ref{fig:ladder}), with a subsequent non-radiative relaxation $\mr{X}^{\bm{-}*}_{\mr{S},\pm3/2}\mathbin{\to}\Xm$ (top solid blue arrow).
	While the temperature dependence of the decay time constant [\subfigref{fig:trpl}{f}] is in this case similar to the one obtained for the low-power excitation, the maximum of \si{\micro}TRPL intensity shifts towards shorter times. The observed unusual temperature dependence of both the decay time and the refiling process suggests that the emission kinetics, in particular the internal $\XXm\!\mathbin{\to}\XmE\!\mathbin{\to}\Xm$ cascade, is strongly temperature dependent, presumably controlled by the acoustic-phonon bath. 

	\begin{figure}[tb]
		\begin{center}
			\includegraphics[width=\columnwidth]{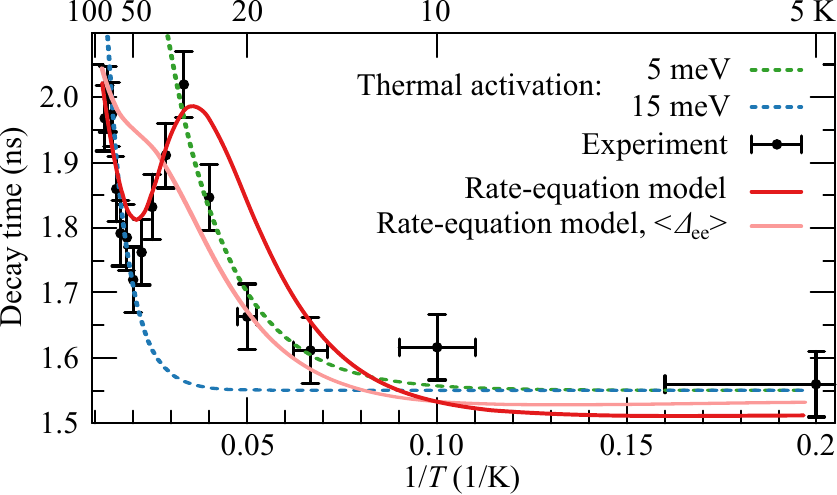}\upcapt
		\end{center}
		\caption{\label{fig:exp_model}Arrhenius plot of PL decay times from the $\Xm$ line (points) along with: two curves for a single thermal activation (dotted lines) and curves resulting form the rate-equation model (solid red lines), the lighter one is the result of averaging over $\varDelta_\mr{ee}$.\upcaptb}
	\end{figure}
	The dependence of lifetimes extracted from the \si{\micro}TRPL data on temperature, replotted in \figref{fig:exp_model} in the Arrhenian form, suggests existence of at least two thermally activated processes. Considering obvious phonon-driven transitions that are present among considered states (solid blue arrows in \figref{fig:ladder}, corresponding to the electron $p\mathbin{\to}s$ relaxation), we attribute the increase of the trion lifetime at $T\mathbin{\approx}\SIrange{60}{80}{\kelvin}$ to the thermally activated excitation from the ground to the excited singlet state. This effectively slows down the opposite process of relaxation, which increases the PL decay time. In fact, the resulting temporal tail of luminescence is formed by a comparatively instantaneous recombination of slowly delivered trion occupation. Also, the thermal quenching of the emission intensity may partially result from such kinetics, as the slowdown of the emission from a given state (into the observed spectral line) results in more emission from other states, which effectively reduces the time-integrated signal. 
	
	Assuming a simple, thermally activated nature of the lifetime increase, we initially estimated the two activation energies to be about 5 and \SI{15}{\milli\electronvolt} (dashed blue and green lines, respectively, in \figref{fig:exp_model}). These values correspond well, to the singlet-triplet energy separation $\varDelta_\mr{ee}$, and the expected value of $s$-$p_y$ splitting $\varDelta_y$\cite{GawelczykPRB2017} in such structures. While the former agrees with our initial guess on the source of the second rise of lifetime, the latter suggests that triplet states are involved in the kinetics that led to the first increase. Additionally, the non-monotonic behavior of lifetime in the moderate temperature range (\SI{\sim 50}{\kelvin}) also indicates that the thermally activated excitation involved here does not happen on the main singlet-singlet relaxation path but rather on another one, possibly through the triplet states. Thus, we attribute the observed feature to an occupation that is stored for a relatively long time in the triplet states (mainly the dark $\mr{X}^{\bm{-}*}_{\mr{T}{,}{\pm}5/2}$) at low temperature. Conversely, at higher $T$ such occupation is thermally released to the excited singlet state, from where it relaxes to the trion ground state, which increases the observed emission lifetime. Such a meta-stable side-path in the relaxation should be disabled when $kT$ overcomes the related activation energy (here $\varDelta_\mr{ee}$), as the rates of relaxation into and excitation from these states equalize. However, for such a mechanism to work, two requirements have to be fulfilled: relaxation from the singlet to triplet states ($\mr{X}^{\bm{-}*}_{\mr{S}}\!\mathbin{\to}\mr{X}^{\bm{-}*}_{\mr{T}}$) has to be relatively fast, and then further relaxation to the trion ground state ($\mr{X}^{\bm{-}*}_{\mr{T}}\!\mathbin{\to}\mr{X}^{\bm{-}}_{\mr{S}}$) as well as triplet recombination ($\mr{X}^{\bm{-}*}_{\mr{T}}\!\mathbin{\to}\mr{e}^{\bm{-}*}$) have to be attenuated. This is obviously not the case for the cascade in its unperturbed form, \textit{i.e.}, without the spin-flip transitions marked with dashed and dotted arrows in \figref{fig:ladder}, since there is no relaxation to triplet states at all.
	
	We utilize a simple rate-equation model to simulate the recombination dynamics in the system: $\dot{p_i}\lr{t}\mathbin{=}\sum_j \Lr{ r_{j\mathbin{\to} i} p_j\lr{t} - r_{i\mathbin{\to} j} p_i\lr{t} }$, where $p_i\lr{t}$ is the occupation of the $i$-th state and $r_{i\mathbin{\to} j}$ are the respective transition rates. For the main relaxation ($\mr{X}^{\bm{-}*}_{\mr{S}}\!\mathbin{\to}\mr{X}^{\bm{-}}_{\mr{S}}$ and $\mr{e}^{\bm{-}*}\!\mathbin{\to}\mr{e}^{\bm{-}}$, both involving the electron $p\mathbin{\to}s$ transition) we assume a fast process associated with the emission of two acoustic phonons\cite{JacakPRB2002,ZibikNatMat2009}. The estimated electron $s$-$p$ splitting corresponds to the energy of two TA phonons from the Brillouin-zone edge, where phonon density of states is enormous. This allows us to set $1/\varGamma\mathbin{=}\SI{100}{\pico\second}$ at $T\mathbin{=}\SI{0}{\kelvin}$. For the radiative recombination we use $1/\gamma\mathbin{=}\SI{1.5}{\nano\second}$. Dashed arrows in \figref{fig:ladder} mark the processes of electron orbital relaxation accompanied by a spin flip,\cite{KhaetskiiPRB2000,GawelczykSST2017} while the dotted arrows depict transitions that involve electron spin flip without orbital relaxation\cite{KhaetskiiPRB2001,Mielnik-PyszczorskiPRB2018}. We are able to obtain a qualitative agreement with the experimental temperature dependence of $\Xm$ lifetimes assuming $1/\varGamma'\mathbin{\ge}\SI{1}{\micro\second}$ for both spin-flip-assisted relaxation processes and very efficient spin flips, $1/\varGamma''\!\mathbin{=} \SI{5}{ns}$. Additional assumption (justified in the case of phonon-driven process due to larger energy separation) that $\mr{X}^{\bm{-}*}_{\mr{S}{,}{\pm}3/2}\!\mathbin{\to}\mr{X}^{\bm{-}*}_{\mr{T}{,}{\pm}5/2}$ is a few times faster (5 times assumed here) than the others leads to improved agreement. The latter is also strongly enhanced when transitions $\varGamma''$ are assumed to origin from some kind of second-order phonon-mediated process characterized by quadratic scaling of the transition rate with the phonon occupation, which leads to a stronger temperature dependence. The solid dark red curve in \figref{fig:exp_model} corresponds to the solution obtained assuming the above-mentioned scenario, which is in a qualitative agreement with the experimental data.
	
	Apart from efficient spin flips also a relatively low singlet-triplet splitting $\varDelta_{\mr{ee}}$ is important for the characteristic dependence to occur. Its low value is actually expected for significant elongation of nanostructures, as it leads to a smaller overlap of electron wave functions. Larger value of $\varDelta_{\mr{ee}}$ in the model reduces the energy separation of the two thermal processes and in consequence eliminates the nonmonotonicity. Thus, we attribute the observed kinetics exclusively to the strong asymmetry of investigated nanostructures causing both enhanced spin relaxation and weakened electron-electron exchange interaction. Due to this, the effect may not be noticed in experiments performed on ensembles of quantum dots that unavoidably contain also more symmetric structures. The solid light red line in \figref{fig:exp_model} results from averaging over the Gaussian distribution of $\varDelta_{\mr{ee}}\mathbin{=}\SI[separate-uncertainty = true]{6.6+-1.25}{\milli\electronvolt}$, which destroys the nonmonotonic behavior. Moreover, the model allows us to predict that $\varDelta_\mr{ee}$ has to be sufficiently different from $\varDelta_{sp}$ for the local minimum of the lifetime to occur, which means that there should be an upper bound for the size of structures that exhibit this feature. 
	Considering the given set of states, we have not found any alternative assumptions that would allow us to explain the observed kinetics. Also the influence of states involving the excited hole has been checked and found to affect the temperature dependence of lifetime of the $\Xm$ line weakly, which justifies neglection of these states. Based on a similar model written for the neutral exciton recombination cascade, we found that such a nontrivial temperature dependence cannot be expected in that case.
	
	In conclusion, we have presented an observation of strong non-monotonic temperature dependence of the negatively charged trion recombination dynamics in highly asymmetric InAs/AlGaInAs/InP QDs. Utilizing the rate-equation model to simulate the carrier-state occupation kinetics in the system, we have qualitatively reproduced the observed dependence. We have found the non-monotonic behavior to result from an additional path of relaxation that leads through the excited trion triplet states. This acts as a meta-stable occupation reservoir that may be thermally emptied (slowdown of emission) and eventually disabled (faster PL decay). To obtain this, we had, however, to assume an efficient electron spin relaxation on a single nanoseconds time scale. Origin of the latter remains an open question that demands decent theoretical investigation.   
	
		The work was supported by Grants No.~{2014/\allowbreak{}14/\allowbreak{}M/\allowbreak{}ST3/\allowbreak{}00821} and No.~{2011/\allowbreak{}02/\allowbreak{}A/\allowbreak{}ST3/\allowbreak{}00152} from the Polish National Science Centre. {\L}.D.\ acknowledges the financial support from the Foundation for Polish Science within the START fellowship. S.H.\ acknowledges support from the State of Bavaria in Germany. We would like to thank Pawe{\l} Machnikowski for valuable discussions and Marcin Syperek for his support in time-resolved measurements. {\L}.D. and M.G. contributed equally to this work.
	
	%
\end{document}